\documentclass[aps,prl,twocolumn,showpacs]{revtex4-1}

\usepackage{bm,graphicx,textcomp,amssymb,amsmath,dcolumn,color}

\newcommand{\ket}[1]{\left|#1\right>}
\newcommand{\bra}[1]{\left<#1\right|}

\newcommand{\nn}{\nonumber\\}

\newcommand{\f}[1]{\mbox{\boldmath$#1$}}
\newcommand{\fk}[1]{\mbox{\boldmath$\scriptstyle#1$}}

\newcommand{\bea}{\begin{eqnarray}}
\newcommand{\ea}{\end{eqnarray}}
\newcommand{\eea}{\end{eqnarray}}
\newcommand{\ord}{\,{\cal O}}

\begin{document}

\title{Environment induced pre-thermalization in the Mott-Hubbard model}

\author{Friedemann Queisser$^{1,2}$ and Ralf~Sch\"utzhold$^{1,2,3}$}

\affiliation{$^1$Fakult\"at f\"ur Physik,
Universit\"at Duisburg-Essen, Lotharstra{\ss}e 1, Duisburg 47057, Germany,}

\affiliation{$^2$Helmholtz-Zentrum Dresden-Rossendorf, 
Bautzner Landstra{\ss}e 400, 01328 Dresden, Germany,}

\affiliation{$^3$Institut f\"ur Theoretische Physik, 
Technische Universit\"at Dresden, 01062 Dresden, Germany.}

\date{\today}

\begin{abstract}
Via the hierarchy of correlations, we study the strongly interacting 
Fermi-Hubbard model in the Mott insulator state and couple it to a Markovian 
environment which constantly monitors the particle numbers 
$\hat n_\mu^\uparrow$ and $\hat n_\mu^\downarrow$ for each lattice site $\mu$. 
As expected, the environment induces an imaginary part $\gamma$ 
(i.e., decay rate) of the 
quasi-particle frequencies $\omega_{\fk{k}}\to\omega_{\fk{k}}-i\gamma$ and 
tends to diminish the correlations between lattice sites. 
Surprisingly, the environment does also steer the state of the system on intermediate 
time scales $\ord(1/\gamma)$ to a pre-thermalized state very similar to a 
quantum quench (i.e., suddenly switching on the hopping rate $J$).
Full thermalization occurs via local on-site heating and takes much longer. 
\end{abstract}


\maketitle

\paragraph{Introduction}

Understanding the quantum dynamics of strongly interacting many-body 
systems is one of the major challenges of contemporary physics. 
Compared to weakly or non-interacting systems, strong interactions can 
induce new and fascinating phenomena. 
One example is the Mott insulator state: 
For a fermionic lattice with a half-filled band, one would expect 
conducting (i.e., metallic) behavior -- but strong interactions 
can make this system insulating \cite{H63,IFT}.

While the ground or thermal equilibrium state of strongly interacting 
systems can already display non-trivial properties, their 
non-equilibrium dynamics can pose even more difficult problems, 
which we are just beginning to understand. 
A conceptually clear and frequently studied example is a quantum quench, 
where one starts in the ground state of a given Hamiltonian
and then suddenly (or non-adiabatically) changes one of the parameters 
of this Hamiltonian.
After that, the initial state will no longer be the ground state in 
general and the time dependence after such a global excitation has 
been studied in various works, see, e.g. \cite{CC06,MWNM07,IC09,SF11,MK08,EKW09,KLA07,BKL10,R09,R10,CDEO08,CFMSE08,FCMSE08,LK08,BRK11,BPBRK12,BPCK12,QKNS14,NS10,NQS14}.

One of the surprises and unexpected results of such non-equilibrium 
dynamics is the phenomenon of pre-thermalization:
Even in systems which are expected to thermalize after a global excitation,
this thermalization dynamics can occur in several stages with different 
time-scales.
Local observables which oscillate on short time scales (after the quench)
approach a quasi-static value on intermediate time-scales -- which is, 
however, different from their thermal value. 
Full thermalization (if it occurs) requires much longer time 
scales. 
As an intuitive picture, pre-thermalization can be understood as 
dephasing of the quasi-particle excitations while full thermalization
requires the exchange of energy and momentum between the 
quasi-particles. 
How strongly interacting quantum many-body systems equilibrate 
is a very important and not fully solved question which has far 
reaching consequences, ranging from solid-state devices such as 
the proposed Mott transistor or other switching processes to 
the observability of quark-gluon plasma or cosmology. 

So far, equilibration and thermalization dynamics after quantum quenches 
and related questions were mostly discussed in closed quantum systems 
undergoing a unitary evolution \cite{MWNM07,IC09,MK08,EKW09,KLA07,BKL10,R09,R10,CDEO08,CFMSE08,FCMSE08,QKNS14}.
However, every real system is always coupled to an environment, 
which can also affect the equilibration and thermalization dynamics. 
In order to fill this gap, we consider a prototypical example for 
a strongly interacting quantum many-body system and study its 
non-equilibrium dynamics after coupling it an environment which is 
assumed to be Markovian. 

\paragraph{The Model}

The lattice system under consideration is described by the Fermi-Hubbard 
Hamiltonian ($\hbar=1$)
\bea
\label{Fermi-Hubbard}
\hat H
=
-\frac{1}{Z}\sum\limits_{\mu,\nu,s}J_{\mu\nu}
\hat c_{\mu,s}^\dagger\hat c_{\nu,s}
+U\sum\limits_\mu\hat n_\mu^\uparrow\hat n_\mu^\downarrow
\,,
\ea
where $\hat c_{\mu,s}^\dagger$ and $\hat c_{\nu,s}$ are the fermionic creation 
and annihilation operators for the spin $s\in\{\uparrow,\downarrow\}$ at the 
lattice sites $\mu$ and $\nu$, respectively.
The corresponding hopping rate is denoted by $J_{\mu\nu}$, where we have 
factored out the coordination number $Z$.
The second term describes the on-site repulsion $U$ with the particle 
number operators $\hat n_\mu^\uparrow$ and $\hat n_\mu^\downarrow$. 
As possible experimental realizations, one could envision fermionic 
atoms in optical lattices \cite{E10,JSGME08,CNLOZ16,BHSON16,PMCJGG16} or electrons in solid state settings \cite{LNW06,DHS03}. 

The above Hamiltonian~\eqref{Fermi-Hubbard} generates the internal unitary 
evolution while the coupling to the Markovian environment is described in 
terms of a master equation with the Lindblad operators $\hat n_{\mu,s}$ 
and the coupling strength $\gamma$
\bea
\label{Lindblad}
\partial_t\hat\rho
=
i\left[\hat\rho,\hat H\right]
+\gamma\sum\limits_{\mu,s}
\left(\hat n_{\mu,s}\,\hat\rho\,\hat n_{\mu,s}-
\left\{\hat n_{\mu,s},\hat\rho\right\}\right)
\,,
\ea
where we have used $\hat n_{\mu,s}^2=\hat n_{\mu,s}$ for fermions. 
Thus, the environment permanently monitors (i.e., weakly measures) the number 
of particles $\hat n_{\mu,s}$ per lattice site $\mu$ for each spin species $s$.
Such an environment could be represented by a bath of bosons which scatter at 
the fermions depending on their position.
For atoms in optical lattices, they could be photons, and for electrons in 
solids, they could be phonons, for example. 

The above master equation~\eqref{Lindblad} can be written in terms of 
Liouville super operators 
\bea
\label{Liouville}
\partial_t\hat\rho
=
\frac{1}{Z}\sum\limits_{\mu,\nu}{\mathfrak L}_{\mu,\nu}\hat\rho
+
\sum\limits_{\mu}{\mathfrak L}_{\mu}\hat\rho
\,,
\ea
where ${\mathfrak L}_{\mu,\nu}$ contains the hopping term $\propto J_{\mu\nu}$
from~\eqref{Fermi-Hubbard} while the on-site interaction term $\propto U$ 
from~\eqref{Fermi-Hubbard} as well as the environment contribution 
$\propto\gamma$ from~\eqref{Lindblad} are encoded in the local contribution 
${\mathfrak L}_{\mu}$.  

\paragraph{Hierarchy of Correlations}

Since the dynamics~\eqref{Lindblad} can only be solved exactly for very small 
lattices (see below), we have to introduce a suitable approximation scheme. 
Here, we employ the hierarchy of correlations \cite{QKNS14,KNQS14,NQS14,QNS12,NQS16,NS10} and consider the reduced 
density matrices $\hat\rho_\mu$ for one site and $\hat\rho_{\mu\nu}$ for two 
sites etc. 
After splitting off the correlations via 
$\hat\rho_{\mu\nu}^{\rm corr}=\hat\rho_{\mu\nu}-\hat\rho_\mu\hat\rho_\nu$ 
and so on, we obtain for the evolution of the on-site density matrix 
\bea
\label{on-site}
\partial_t\hat\rho_\mu
&=&
\frac{1}{Z}\sum\limits_{\nu}
{\rm Tr}_\nu\left\{{\mathfrak L}_{\mu,\nu}
\hat\rho_\mu\hat\rho_\nu+{\mathfrak L}_{\mu,\nu}\hat\rho_{\mu\nu}^{\rm corr}
\right\}
+{\mathfrak L}_{\mu}\hat\rho_\mu
\nn
&=&
f_1(\hat\rho_\nu,\hat\rho_{\mu\nu}^{\rm corr})
\,.
\ea
In analogy, the time evolution of the two-site correlations can be derived 
from~\eqref{Liouville} and also depends on the on-site density matrices as 
well as the three-site correlators 
\bea
\label{two-site}
\partial_t\hat\rho_{\mu\nu}^{\rm corr}
=
f_2(
\hat\rho_\nu,\hat\rho_{\mu\nu}^{\rm corr},\hat\rho_{\mu\nu\sigma}^{\rm corr})
\,.
\ea
In order to truncate this infinite set of recursive equations, we exploit the 
hierarchy of correlations in the formal limit of large coordination numbers 
$Z\gg1$.
With completely the same arguments as in \cite{NS10}, it can be shown that the 
two-site correlations are suppressed via $\hat\rho_{\mu\nu}^{\rm corr}=\ord(1/Z)$
in comparison to the on-site density matrix $\hat\rho_{\mu}=\ord(Z^0)$ 
and the three-site correlators even stronger via 
$\hat\rho_{\mu\nu\sigma}^{\rm corr}=\ord(1/Z^2)$, and so on.
Note that the derivation in \cite{NS10} works in completely the same way here 
because the environment acts locally, i.e., on each lattice site separately,
and thus only changes the local Liovillian ${\mathfrak L}_{\mu}$ 
in~\eqref{Liouville}. 

This hierarchy of correlations facilitates the following iterative 
approximation scheme:
To zeroth order in $1/Z$, we may approximate~\eqref{on-site} via 
$\partial_t\hat\rho_\mu\approx f_1(\hat\rho_\nu,0)$ which yields the mean-field 
solution $\hat\rho_\mu^0$.
As the next step, we may insert this solution $\hat\rho_\mu^0$ 
into~\eqref{two-site} and obtain to first order in $1/Z$ the following 
approximate set of linear and inhomogeneous equations for the correlations
\bea
\label{two-site-linear}
\partial_t\hat\rho_{\mu\nu}^{\rm corr}
\approx
f_2(\hat\rho_\nu^0,\hat\rho_{\mu\nu}^{\rm corr},0)
\,.
\ea
The solution of this set of equations describes the propagation 
(and damping) of the quasi-particles, insertion back into~\eqref{on-site} 
then yields their back-reaction onto the mean field.  

\paragraph{Mean-field Ansatz}

Let us study the propagation (and damping) of the quasi-particles according 
to~\eqref{two-site-linear} for a concrete example. 
For the mean-field solution $\hat\rho_\mu^0$, we assume a homogeneous and 
spin-symmetric (i.e., unpolarized) state, which can be described by the 
general ansatz 
\bea
\label{ansatz} 
\hat\rho_\mu^0=
p_0\ket{0}\bra{0}+
p_1\left(\ket{\uparrow}\bra{\uparrow}+\ket{\downarrow}\bra{\downarrow}\right)+
p_2\ket{\uparrow\downarrow}\bra{\uparrow\downarrow}
\,,
\ea
with the probabilities for zero $p_0$, one $p_1$, and two particles $p_2$
on the lattice site $\mu$.
For the Fermi-Hubbard Hamiltonian~\eqref{Fermi-Hubbard} and the 
Lindblad operators $\hat n_{\mu,s}$ in~\eqref{Lindblad}, this 
ansatz automatically satisfies the zeroth-order (mean-field) equation 
$\partial_t\hat\rho_\mu^0=f_1(\hat\rho_\mu^0,0)$.

Since we want to include the Mott insulator state \cite{SDM00,IFT}, we assume half filling 
$\langle\hat n_\mu^\uparrow+\hat n_\mu^\downarrow\rangle=2p_1+2p_2=1$.
Together with the normalization $p_0+2p_1+p_2=1$, this fixes all probabilities 
except one, which can be parametrized by the double occupancy 
$\mathfrak D=\langle\hat n_\mu^\uparrow\hat n_\mu^\downarrow\rangle=p_2$. 
It vanishes in the Mott insulator state $p_0=p_2=0$, 
but in the infinite-temperature limit $p_0=p_1=p_2$, 
it tends to 1/4.

Note that, since the ansatz~\eqref{ansatz} obeys the zeroth-order (mean-field) 
equation $\partial_t\hat\rho_\mu^0=f_1(\hat\rho_\mu^0,0)$, the double occupancy 
$\mathfrak D$ is constant to lowest order (in $1/Z$).
However, including the back-reaction of the quasi-particles and their quantum 
or thermal fluctuations onto the mean field, it will change in general (see below).

\paragraph{Quasi-particles}

Inserting the ansatz~\eqref{ansatz} into the equation~\eqref{two-site-linear}
for the correlations, we find the following set of relevant correlation functions,
see also \cite{QKNS14}
\bea
f^{00}_{\mu\nu,s}
&=&
\langle\hat c_{\mu,s}^\dagger(1-\hat n_{\mu,\bar s})\,
\hat c_{\nu,s}(1-\hat n_{\nu,\bar s})
\rangle
\nn
f^{01}_{\mu\nu,s}
&=&
\langle\hat c_{\mu,s}^\dagger(1-\hat n_{\mu,\bar s})\,
\hat c_{\nu,s}\hat n_{\nu,\bar s}
\rangle
\nn
f^{10}_{\mu\nu,s}
&=&
\langle\hat c_{\mu,s}^\dagger\hat n_{\mu,\bar s}\,
\hat c_{\nu,s}(1-\hat n_{\nu,\bar s})
\rangle
\nn
f^{11}_{\mu\nu,s}
&=&
\langle\hat c_{\mu,s}^\dagger\hat n_{\mu,\bar s}\,
\hat c_{\nu,s}\hat n_{\nu,\bar s}
\rangle
\,,
\ea
with $s\in\{\uparrow,\downarrow\}$ denoting the spin and $\bar s$ 
the opposite spin.
All other correlators vanish to first order (in $1/Z$). 

As we obtain the same dynamics for both spin species $s$, 
we omit the spin index $s$ in the following.  
Assuming spatial homogeneity, we Fourier transform the 
above correlation functions and \eqref{two-site-linear} 
becomes  
\bea
\label{time-f_k}
\left(i\partial_t+i\gamma\right)f^{00}_{\fk{k}}
&=&
\frac{J_{\fk{k}}}{2}\left(f^{10}_{\fk{k}}-f^{01}_{\fk{k}}\right) 
=
-\left(i\partial_t+i\gamma\right)f^{11}_{\fk{k}}
\nn
\left(i\partial_t-U+i\gamma\right)f^{01}_{\fk{k}}
&=&
\frac{J_{\fk{k}}}{2}\left(f^{11}_{\fk{k}}-f^{00}_{\fk{k}}\right) 
-\frac{J_{\fk{k}}}{4}\left(1-4{\mathfrak D}\right) 
\nn
&=&
-\left(i\partial_t+U+i\gamma\right)f^{10}_{\fk{k}}
\,.
\ea
For time-independent parameters $\gamma$, $J_{\fk{k}}$, $U$, 
and $\mathfrak D$, we may diagonalize the above linear system of 
equations and thereby obtain four eigen-frequencies, 
two of them read 
\bea
\label{eigen-frequencies}
\omega^\pm_{\fk{k}}=\pm\sqrt{U^2+J_{\fk{k}}^2}-i\gamma
\ea
while the other two simply are $\omega^0_{\fk{k}}=-i\gamma$. 
We see that all eigen-frequencies acquire the same imaginary part $-i\gamma$ 
which just corresponds to an exponential decay $e^{-\gamma t}$.  
This describes the damping of the quasi-particles induced by the 
coupling to the environment. 

\paragraph{Pre-thermalization}

Due to this  exponential decay $e^{-\gamma t}$, the correlation functions 
approach the following asymptotic state 
(again assuming that $\mathfrak D$ is constant) 
\bea
\label{asymptotic}
f^{00}_{\fk{k},\rm asy} 
&=&
-\frac{J_{\fk{k}}^2}{U^2+J_{\fk{k}}^2+\gamma^2}\,
\frac{1-4{\mathfrak D}}{4}
=
-f^{11}_{\fk{k},\rm asy} 
\,,
\nn
f^{01}_{\fk{k},\rm asy} 
&=&
-\frac{J_{\fk{k}}\left(U+i\gamma\right)}{U^2+J_{\fk{k}}^2+\gamma^2}\,
\frac{1-4{\mathfrak D}}{4}
=
\left(f^{10}_{\fk{k},\rm asy}\right)^*
\,,
\ea
which is independent of the initial state, i.e., the initial values  
$f^{00}_{\fk{k}}(t=0)$, $f^{01}_{\fk{k}}(t=0)$, $f^{10}_{\fk{k}}(t=0)$, and 
$f^{11}_{\fk{k}}(t=0)$.
As one would expect, the correlations are suppressed for large $\gamma$ 
and go to zero in the limit $\gamma\to\infty$.

However, it is interesting to note that this asymptotic 
state~\eqref{asymptotic} is different from the ground state,
even for $\gamma=0$ and ${\mathfrak D}=0$, where we have 
\bea
f^{00}_{\fk{k},\rm ground} 
=
\frac14\left(\frac{U}{\sqrt{U^2+J_{\fk{k}}^2}}-1\right)
=
-f^{11}_{\fk{k},\rm ground} 
\,.
\ea
Already for small $J_{\fk{k}}$, we observe a factor of two difference to the 
asymptotic state~\eqref{asymptotic}, see also \cite{MK09,QKNS14}. 

In fact, in the limit of small $\gamma$, the above asymptotic state~\eqref{asymptotic}
coincides with the pre-thermalized state after a quantum quench, where one starts 
in the ground state with $J=0$ and then suddenly switches on $J$ to its final value,
see, e.g., \cite{QKNS14}. 
This coincidence seems to be a rather general property.
To understand why, let us write the linear system of 
equations~\eqref{time-f_k} in matrix form 
\bea
\label{matrix-form}
\partial_t\f{f}_{\fk{k}}
=
\f{M}_{\fk{k}}\cdot\f{f}_{\fk{k}}
-\gamma\f{f}_{\fk{k}}
+\f{s}_{\fk{k}}
\,,
\ea
with a time-idependent matrix $\f{M}_{\fk{k}}$ describing the Hamiltonian 
evolution, i.e., depending on $J_{\fk{k}}$ and $U$.
Neglecting back-reaction, i.e., assuming that the double occupancy 
$\mathfrak D=\langle\hat n_\mu^\uparrow\hat n_\mu^\downarrow\rangle$
is time-independent, the source term $\f{s}_{\fk{k}}$ is also constant.
Then, due to the damping term $\gamma$, the correlations approach the 
asymptotic state 
\bea
\label{asymptotic-matrix}
\f{f}_{\fk{k}}^{\rm asy}
=
\left(\gamma\f{1}-\f{M}_{\fk{k}}\right)^{-1}\cdot\f{s}_{\fk{k}}
\,.
\ea
Now, the limit $\gamma\to0$ could be problematic if the source term 
$\f{s}_{\fk{k}}$ would have contributions in the kernel 
${\rm ker}(\f{M}_{\fk{k}})$ of the matrix $\f{M}_{\fk{k}}$, i.e., 
the sub-space of zero eigenvalue.
In this case, the linear evolution according to~\eqref{matrix-form} 
without environment $\gamma=0$ would imply linearly growing modes -- 
which indicate an instability (e.g., if the mean-field ansatz 
$\hat\rho_\mu^0$ does not describe a stationary state).
In the various scenarios investigated by us (see \cite{QKNS14}), 
we did not encounter 
this problem and hence we assume $\f{s}_{\fk{k}}\perp{\rm ker}(\f{M}_{\fk{k}})$ 
in the following and omit the kernel of the matrix $\f{M}_{\fk{k}}$.

In the sub-space orthogonal to the kernel ${\rm ker}(\f{M}_{\fk{k}})$,
we may invert the matrix $\f{M}_{\fk{k}}$ and the limit $\gamma\to0$
of the asymptotic state~\eqref{asymptotic-matrix} reads 
$\f{f}_{\fk{k}}^{\rm asy}=-\f{M}_{\fk{k}}^{-1}\cdot\f{s}_{\fk{k}}$.  
Now let us compare this state to the pre-thermalized state after a 
quantum quench (without environment). 
If we start initially in the ground state for $J=0$,
we have vanishing correlations initially $\f{f}_{\fk{k}}(t=0)=0$.
At time $t=0$, we switch on the hopping rate $J$.
The time evolution afterwards can be obtained by 
solving~\eqref{matrix-form} for $\gamma=0$ and 
vanishing initial correlations, which yields  
\bea
\label{quench}
\f{f}_{\fk{k}}(t)
=
\left(\exp\left\{\f{M}_{\fk{k}}t\right\}-\f{1}\right)
\cdot
\f{M}_{\fk{k}}^{-1}\cdot\f{s}_{\fk{k}}
\,.
\ea
As a result, the Fourier modes $\f{f}_{\fk{k}}(t)$ of the correlations 
oscillate with the (non-zero) eigen-frequencies of the matrix $\f{M}_{\fk{k}}$.
The Fourier transformation back to position space then involves a sum over 
many Fourier modes with different oscillating phases, which gives the 
usual pre-thermalization dynamics as in Fig...
The long-time limit then corresponds to the time average 
$\overline{\f{f}_{\fk{k}}}$ where the oscillating exponentials cancel 
$\overline{\f{f}_{\fk{k}}}=-\f{M}_{\fk{k}}^{-1}\cdot\f{s}_{\fk{k}}$. 
Hence, the coincidence of the pre-thermalized state (after a quench) 
and the $\gamma\to0$ limit of the asymptotic state with environment 
seems to be a general phenomenon -- as long as arguments along the 
lines explained above apply. 

Note that the simple matrix form~\eqref{matrix-form} applies to cases 
where all correlations are damped at the same rate $\gamma$.
While this is true for the system under investigation, cf.~\eqref{time-f_k}, 
one might have different damping rates $\gamma_{1,2,\dots}$ for other scenarios.  
However, this just amounts to replacing $\gamma\f{1}$ by a different matrix 
(assumed to be positive definite) while the rest of the arguments applies 
in the same way. 

%
%
%
%

\begin{figure}
\includegraphics{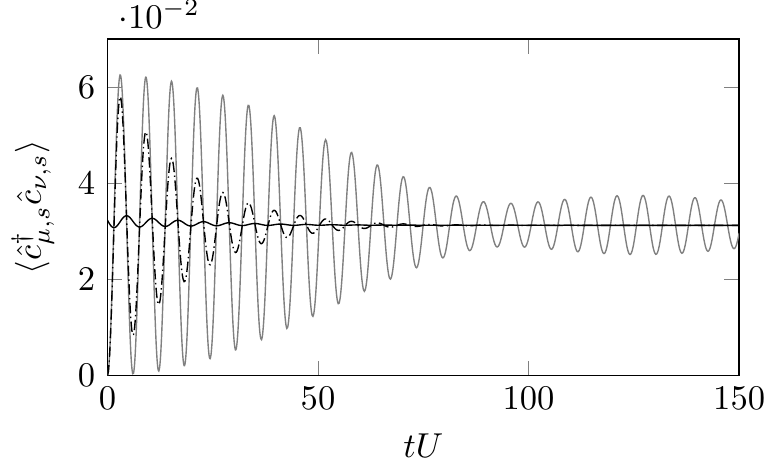}
\includegraphics{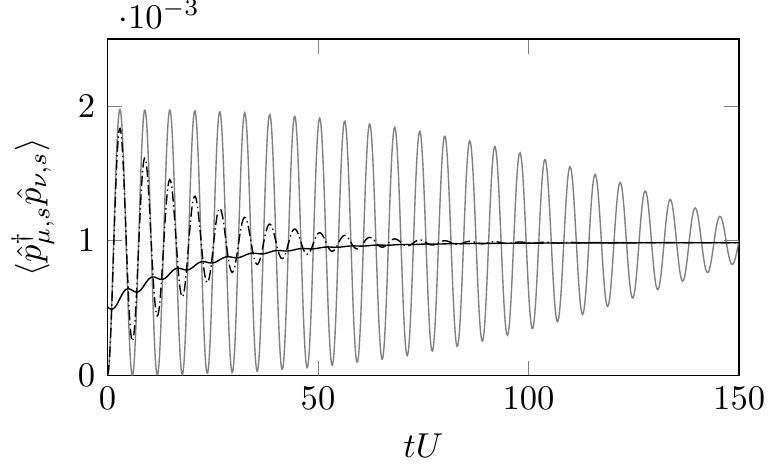}
\caption{Top: $\langle \hat{c}^\dagger_{\mu,s} \hat{c}_{\nu,s}\rangle$-correlations for next neighbours. Bottom:
$\langle \hat{p}^\dagger_{\mu,s} \hat{p}_{\nu,s}\rangle$-correlations for next but one neighbours. Gray curves: Quench 
from $J/U=0$ to $J/U=0.4$ with $\gamma=0$, Dashed curves: Quench 
from $J/U =\gamma/U=0$ to $J/U=0.4$ and $\gamma/U=0.05$, solid black curves: Evolution from the ground state
to the prethermalized state after coupling the system to the environment with $\gamma/U=0.05$.}\label{Fig1}
\end{figure}

\begin{figure}
\includegraphics{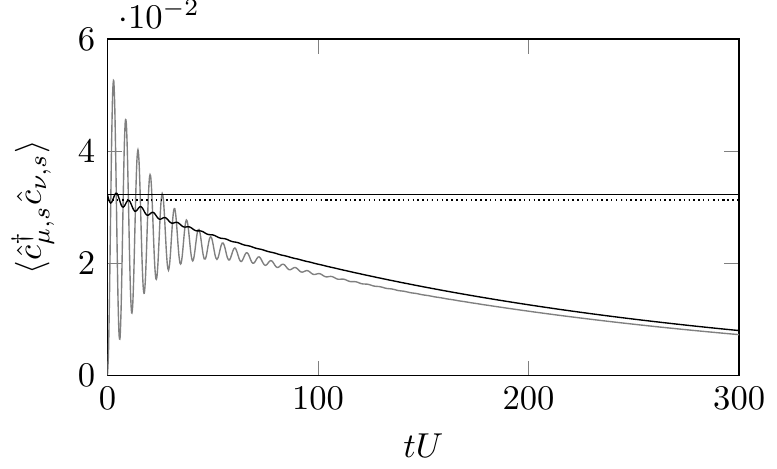}
\includegraphics{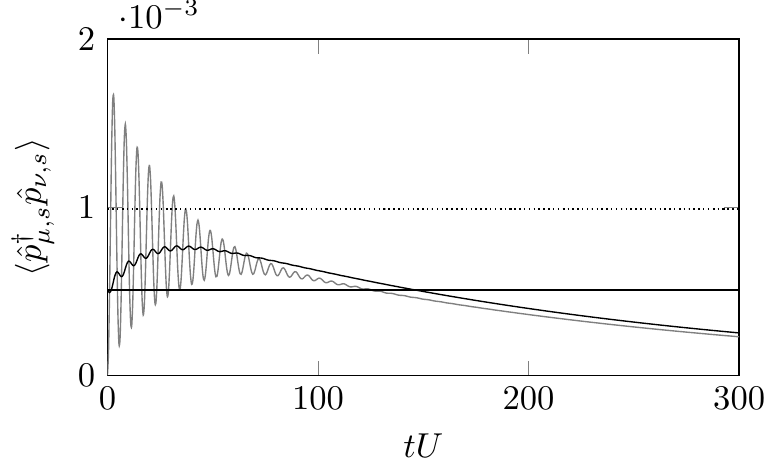}
\caption{Top: $\langle \hat{c}^\dagger_{\mu,s} \hat{c}_{\nu,s}\rangle$-correlations for next neighbours with 
backreaction. Bottom:
$\langle \hat{p}^\dagger_{\mu,s} \hat{p}_{\nu,s}\rangle$-correlations for next but one neighbours with backreaction.
Gray curves: Quench from $J/U =\gamma/U=0$ to $J/U=0.4$ and $\gamma/U=0.05$, solid black curves: Evolution from the ground state
after coupling the system to the environment with $\gamma/U=0.05$.
Solid line: Ground state, dotted line: prethermalized state.}
\end{figure}

%
%
%
%
%
%
%

\paragraph{Back-reaction}

So far, we have neglected the back-reaction of the quantum or thermal 
fluctuations of the quasi-particles onto the mean field and assumed 
that $\hat\rho_\mu^0$ and thus the double occupancy $\mathfrak D$ 
are constant. 
In order to study this back-reaction, we insert the 
(generally time-dependent) solutions 
$f^{00}_{\fk{k}}$, $f^{01}_{\fk{k}}$, $f^{10}_{\fk{k}}$, and $f^{11}_{\fk{k}}$ 
back into into~\eqref{on-site} which gives 
\bea
\label{back-reaction}
i\partial_t{\mathfrak D}=
\frac{2}{N}\sum\limits_{\fk{k}}J_{\fk{k}}
\left(f^{01}_{\fk{k}}-f^{10}_{\fk{k}}
\right)
\,.
\ea
We see that even the asymptotic state~\eqref{asymptotic} 
can induce a change of $\mathfrak D$ provided that $\gamma\neq0$.
For example, starting in the Mott insulator phase with 
zero or small $\mathfrak D$, it would slowly grow due 
to local on-site heating induced by the coupling to the 
environment. 

However, this growth rate is much slower than the damping of 
correlations and quasi-particles with their decay rate $\gamma$.
From the above equation~\eqref{back-reaction}, we may estimate 
that this local on-site heating occurs on much longer time scales 
\bea
\label{longer-time-scales} 
\tau_{\rm thermal}\sim\frac{U^2}{J^2}\,\frac{Z}{\gamma}\gg\frac1\gamma
=\tau_{\rm decay}
\,,
\ea
where the factor of $Z$ stems from the Fourier transform 
(assuming an isotropic lattice). 

For late times $t\gg\tau_{\rm thermal}$, the double occupancy tends 
to 1/4 and thus all correlations between lattice sites vanish, as 
we may infer from~\eqref{asymptotic}. 
This final state $\hat\rho_\mu\propto\f{1}$ corresponds to the 
infinite-temperature limit, which is consistent with the fact 
that the considered Markovian environment acts as an infinite 
temperature heat bath.  

\paragraph{Hubbard Dimer}

The previous investigations were based on the hierarchy of correlations,
which can be motivated by the formal limit $Z\to\infty$.
However, the results obtained via this method (such as the damping 
of quasi-particles and pre-thermalization) can be applied 
qualitatively to scenarios beyond this approximation.
To see this, let us consider a simple system which admits an 
exact solution -- the Fermi-Hubbard model consisting of two lattice 
sites (Hubbard dimer) with one spin-up and one spin-down particle.

To simplify the analysis further, we consider states which are fully 
symmetric with respect to a permutation 
of the lattice sites 
$\mu=1$ and $\nu=2$ and are invariant under spin-flips (i.e., unpolarized).
Again, the on-site matrices $\hat\rho_1^0=\hat\rho_2^0$ can be fully 
parametrized by the double occupancy $\mathfrak D$ via the ansatz~\eqref{ansatz}.
Furthermore, as we only have one particle per spin species, 
the particle-particle $f^{11}_{12,s}$ and hole-hole $f^{00}_{12,s}$ 
correlators vanish and only the particle-hole correlators $f^{01}_{12,s}$ and 
$f^{10}_{12,s}$ remain. 
Finally, the only remaining non-zero expectation values are two 
higher-order correlators 
$\langle\hat c_{1\uparrow}^\dagger\hat c_{1\downarrow}
\hat c_{2\downarrow}^\dagger\hat c_{2\uparrow}\rangle$
and
$\langle\hat c_{1\uparrow}^\dagger\hat c_{1\downarrow}^\dagger 
\hat c_{2\downarrow}\hat c_{2\uparrow}\rangle$.

Now, combining these five relevant expectation values 
(in the mentioned order) into a vector $\f{v}$, 
the time-evolution can be described exactly by a $5\times5$-matrix 
\bea
\label{5x5}
i\partial_t\f{v}
&=&
\left(
\begin{array}{ccccc}
 0 & -J & J & 0 & 0 \\
 -4J & -U-i\gamma & 0 & -2J & J \\
 4J & 0 & +U-i\gamma & 2J & -J \\
 0 & -J & J & -2i\gamma & 0 \\
 0 & 2J & -2J & 0 & -4i\gamma
\end{array}
\right)
\cdot\f{v}
\nn
&&+\f{s}
\,,
\ea
where $\f{s}=(0,J,-J,0,0)^{\rm T}$ are the source terms already discussed above.
This simple matrix equation describes the exact evolution -- 
but, unfortunately, there is no simple closed expression for the 
eigenvalues and eigenvectors of this matrix. 
Thus, let us discuss some limiting cases: 
Without the environment, i.e., for $\gamma=0$, two eigenvalues correspond to 
the quasi-particle energy $\pm\sqrt{16J^2+U^2}$ while the other three vanish.
In order to include the environment, we consider the strongly interacting 
regime where $J\ll U$, i.e., deep in the Mott insulator phase. 
In this regime, one may infer from the diagonal of the above matrix~\eqref{5x5}
that four eigenvalues have imaginary parts of order $\gamma$, which corresponds 
to the decay of the four correlation functions to a pre-thermalized state 
on a time scale of order $1/\gamma$.
However, the remaining eigenvalue (which corresponds to the evolution 
of the double occupancy $\mathfrak D$) is much smaller 
\bea
\lambda=-\frac{8i\gamma J^2}{U^2+\gamma^2}+\ord\left(\frac{J^3}{U^3}\right)
\Rightarrow |\lambda|\ll \gamma
\,.
\ea
We see that, while all the correlations approach their pre-thermalized state 
on a time scale of order $1/\gamma$, full thermalization is governed by the 
above eigenvalue and thus occurs on much longer times. 
Again, due to local on-site heating, the state $\f{v}$ approaches the final 
state $\f{v}=(1/4,0,0,0,0)^{\rm T}$, which corresponds to an 
infinite-temperature state $\hat\rho_{12}\propto\f{1}$. 

\paragraph{Conlcusions}

As a prototypical example for an interacting quantum many-body system, 
we consider the Fermi-Hubbard model~\eqref{Fermi-Hubbard} and couple it 
to a Markovian environment~\eqref{Lindblad} which permanently performs 
weak measurements of the particle numbers $\hat n_\mu^\uparrow$ and 
$\hat n_\mu^\downarrow$ for each lattice site $\mu$. 
Via the hierarchy of correlations, we derive the evolution 
equations~\eqref{time-f_k} for the correlations, 
which are linear to first order (in $1/Z$), as well as 
their back-reaction~\eqref{back-reaction} onto the mean field. 

As expected, the coupling $\gamma$ to the environment induces an imaginary 
part of the eigen-frequencies~\eqref{eigen-frequencies} leading to 
a decay of the quasi-particles and tends to suppress the correlations.
Quite surprisingly, this damping mechanism does also induce the 
phenomenon of pre-thermalization quite analogous to a quantum quench.
For small $\gamma$, the correlations even approach the same 
pre-thermalized state as after a quench. 
As our general arguments from \eqref{matrix-form} to \eqref{quench} indicate, 
this seems to be a general phenomena and shows that the 
environment induced damping of quasi-particles has a very similar effect 
as the dephasing of quasi-particles after a quench. 

Taking the back-reaction~\eqref{back-reaction} into account, we find that 
the system eventually approaches a thermal state of infinite temperature.
However, this on-site heating process is much slower and requires time 
scales~\eqref{longer-time-scales} much longer than the intermediate 
time scale $\ord(1/\gamma)$ of pre-thermalization. 
Finally, in order to test the reliability of our approximation scheme, 
we considered the exactly solvable case of the two-site Fermi-Hubard model 
(Hubbard dimer) and found qualitatively the same results.
We also considered the Mott-Neel state displaying anti-ferromagnetic spin 
ordering (see the supplement) and found analogous behavior. 

\acknowledgments 

\paragraph{Acknowledgements}

The authors acknowledge support by DFG (grant SFB-1242, projects B03 and B07).

\section{Hierarchy of Correlations}

Up to first order in $1/Z$, the equations of motion for the double occupancy and the two-point-correlations
have the explicit form 
\begin{align}
i\partial_t \mathfrak{D}& =
\frac{1}{Z}\sum_{\kappa,s}J_{\mu\kappa}(f^{01}_{\kappa\mu, s}-f^{10}_{\mu\kappa,s})+\mathcal{O}(1/Z^2)\label{onsite}\\
i\partial_t f^{ij}_{\mu\nu,s}&=\frac{1}{Z}\sum_{\kappa,l}J_{\mu\kappa}\langle \hat{n}^i_{\mu,\bar{s}}\rangle
f^{lj}_{\kappa\nu, s}
-\frac{1}{Z}\sum_{\kappa,l}J_{\nu\kappa}\langle \hat{n}^j_{\nu,\bar{s}}\rangle
f^{il}_{\mu\kappa,s}\nonumber\\
&+\frac{J_{\mu\nu}}{Z}[\langle \hat{n}^i_{\mu,\bar{s}}\rangle\langle \hat{n}^1_{\nu,s}\hat{n}^j_{\nu,\bar{s}}\rangle-
\langle \hat{n}^j_{\nu,\bar{s}}\rangle\langle \hat{n}^1_{\mu, s}\hat{n}^i_{\mu,\bar{s}}\rangle]\nonumber\\
&-(U^i-U^j+i\gamma)f^{ij}_{\mu\nu,s}+\mathcal{O}(1/Z^2)\,,\label{twosite}
\end{align}
where we used the shorthand notation $\hat{n}^1_{\mu, s}=\hat{n}_{\mu, s}$ and $\hat{n}^0_{\mu, s}=1-\hat{n}_{\mu, s}$.
After a Fourier transformation of (\ref{twosite}) for a spatially homogeneous system at half filling, one obtains the set of 
equations (9) in the letter.

Since the hierarchical set of equations (\ref{twosite}) is derived in real space, we are not 
restricted to spatially homogeneous systems.
For example, the fermionic Hubbard system in a hypercubic lattice prefers to be in a staggered Mott-N\'eel state
with sublattices $A$ and $B$ if the temperature is sufficiently low.
Assuming to lowest order a perfect staggering, $\hat{n}^A_s=\hat{n}^B_{\bar{s}}=1$, the Fourier components 
of the correlation functions satisfy the equations
\begin{align}\label{staggered}
(i\partial_t+i\gamma)f^{00,AA}_{\mathbf{k},s}&=J_\mathbf{k}(f^{10,BA}_{\mathbf{k},s}-f^{01,AB}_{\mathbf{k},s})\\
(i\partial_t+i\gamma)f^{11,BB}_{\mathbf{k},s}&=J_\mathbf{k}(f^{01,AB}_{\mathbf{k},s}-f^{10,BA}_{\mathbf{k},s})\\
(i\partial_t-U+i\gamma )f^{01,AB}_{\mathbf{k},s}&=J_\mathbf{k}(f^{11,BB}_{\mathbf{k},s}-f^{00,AA}_{\mathbf{k},s})
-J_\mathbf{k}\\
(i\partial_t +U+i\gamma)f^{10,BA}_{\mathbf{k},s}&=J_\mathbf{k}(f^{00,AA}_{\mathbf{k},s}-f^{11,BB}_{\mathbf{k},s})
+J_\mathbf{k}\,.
\end{align}
After a quantum quench, the correlation functions approach the asymptotic state
\begin{align}
f^{11,BB}_{\mathbf{k},s}&=-f^{00,AA}_{\mathbf{k},s}=\frac{2J_\mathbf{k}^2}{\gamma^2+4J_\mathbf{k}^2+U^2}\\
f^{01,AB}_{\mathbf{k},s}&=\left(f^{10,BA}_{\mathbf{k},s}\right)^*=\frac{J_\mathbf{k}(U+i\gamma)}{\gamma^2+4J_\mathbf{k}^2+U^2}
\end{align}
whereas the ground state correlations ($\gamma=0$) have the form
\begin{align}
f^{11,BB}_{\mathbf{k},s}&=-f^{00,AA}_{\mathbf{k},s}=\frac{1}{2}\left(1-\frac{U}{\sqrt{4 J_\mathbf{k}^2+U^2}}\right)\\
f^{01,AB}_{\mathbf{k},s}&= f^{10,BA}_{\mathbf{k},s}=\frac{J_\mathbf{k}}{\sqrt{4 J_\mathbf{k}^2+U^2}}\,.
\end{align}
\section{Hubbard Dimer}
In general, the dynamics of spins on two lattice sites can be described with a set of 16 coupled equations.
Under the assumption that the state is symmetrix w.r.t. spin- and lattice-permutation, the system reduces 
to 5 coupled equations.
With the definitions
\begin{align}
\mathfrak{D}&=\langle \hat{n}_{1,s}\hat{n}_{1,\bar s}\rangle
=\langle \hat{n}_{2,s}\hat{n}_{2,\bar s}\rangle\\
\mathfrak{F}&=\langle \hat{c}^\dagger_{1,s}\hat{n}_{1,\bar s}\hat{c}_{2,s}(1-\hat{n}_{2,\bar s})\rangle
+\langle \hat{c}^\dagger_{2,s}\hat{n}_{2,\bar s}\hat{c}_{1,s}(1- \hat{n}_{1,\bar s})\rangle\\
%
\mathfrak{S}&=\langle \hat{c}^\dagger_{1,s}\hat{c}_{1,\bar s}\hat{c}^\dagger_{2,\bar s}\hat{c}_{2,s}\rangle\\
\mathfrak{H}&=\langle c^\dagger_{1\uparrow}c^\dagger_{1\downarrow}c_{2\uparrow}c_{2\downarrow}\rangle
+\langle c^\dagger_{2\uparrow}c^\dagger_{2\downarrow}c_{1\uparrow}c_{1\downarrow}\rangle
\end{align}
we can write the dynamics of the Dimer as
\begin{align}
i\partial_t \mathfrak{D}&=J(\mathfrak{F}^*-\mathfrak{F})\\
(i\partial_t+U+i\gamma)\mathfrak{F}&=-4 J \mathfrak{D}-2 J \mathfrak{S}+J\mathfrak{H}+J\\
(i\partial_t-U+i\gamma)\mathfrak{F}^*&=4 J \mathfrak{D}+2 J \mathfrak{S}-J\mathfrak{H}-J\\
(i\partial_t+2i\gamma)\mathfrak{S}&=J(\mathfrak{F}^*-\mathfrak{F})\\
(i\partial_t+4i\gamma)\mathfrak{H}&=-2J(\mathfrak{F}^*-\mathfrak{F})
\end{align}
For a finite damping rate, the system runs into the infinite temperature 
state with $\mathfrak{D}=1/4$ and $\mathfrak{F}=\mathfrak{S}=\mathfrak{H}=0$.
For $\gamma=0$, the prethermalized state of the system is given by
\begin{align}
\mathfrak{D}_\mathrm{asy}&=\mathfrak{S}_\mathrm{asy}=-\frac{\mathfrak{H}_\mathrm{asy}}{2}=\frac{2 J^2}{16J^2+U^2}\\
\mathfrak{F}_\mathrm{asy}&=\mathfrak{F}^*_\mathrm{asy}=\frac{JU}{16J^2+U^2}\,.
\end{align}
The ground state of the system can be determined by an adiabatic increase of the 
hopping rate to a finite value:
\begin{align}
\mathfrak{D}_\mathrm{ground}&=\mathfrak{S}_\mathrm{ground}=-\frac{\mathfrak{H}_\mathrm{ground}}{2}\nonumber\\
&=
\frac{1}{8}\left(1-\frac{U}{\sqrt{16 J^2+U^2}}\right)\\
\mathfrak{F}_\mathrm{ground}&=\mathfrak{F}^*_\mathrm{ground}=\frac{J}{\sqrt{16J^2+U^2}}
\end{align}


\begin{thebibliography}{99}
 
\bibitem{H63} J.~Hubbard, Proc. R. Soc. Lond. A {\bf 276}, 238 (1963).
\bibitem{IFT} M.~Imada, A.~Fujimori, and Y.~Tokura, Rev.~Mod.~Phys. {\bf 70} (1998).
%
\bibitem{CC06} P.~Calabrese and J.~Cardy, Phys.~Rev.~Lett. {\bf 96}, 136801 (2006).
\bibitem{MWNM07} S.~R. Manmana, S.~Wessel, R.~M.~Noack, and A.~Muramatsu,
Phys.~Rev.~Lett. {\bf 98}, 210405 (2007).
\bibitem{IC09} A.~Iucci and M.~A.~Cazalilla, Phys. Rev. A {\bf 80}, 063619 (2009).
\bibitem{SF11} M.~Schir\'o and M.~Fabrizio, Phys.~Rev.~B {\bf 83}, 165105 (2011).
\bibitem{MK08} M.~Moeckel and S.~Kehrein, Phys.~Rev.~Lett. {\bf 100}, 175702 (2008).
\bibitem{EKW09} M.~Eckstein, M.~Kollar, and P.~Werner, Phys.~Rev.~Lett. {\bf 103}, 056403 (2009).
%
%
%
\bibitem{KLA07} C.~Kollath, A.~M.~L \"auchli, and E.~Altman, Phys.~Rev.~Lett.
{\bf 98}, 180601 (2007).
\bibitem{BKL10} G.~ Biroli, C.~Kollath, and  A.~M.~L\"auchli, Phys.~Rev.~Lett.
{\bf 105}, 250401 (2010).
\bibitem{R09} G.~Roux, Phys.~Rev.~A {\bf 79}, 021608 (2009).
\bibitem{R10} G.~Roux, Phys.~Rev.~A {\bf 81}, 053604 (2010).
\bibitem{CDEO08} M.~Cramer, C.~M.~Dawson, J.~Eisert, and T.~J.~Osborne, Phys.~Rev.~Lett.
{\bf 100}, 030602 (2008).
\bibitem{CFMSE08} M.~Cramer , A.~Flesch, I.~P.~McCulloch, U.~Schollw \"ock, and J. Eisert, 
Phys.~Rev.~Lett. {\bf 101}, 063001 (2008).
\bibitem{FCMSE08} A.~Flesch, M.~Cramer, I.~P. McCulloch, U.~Schollw\"ock, and J.~Eisert, 
Phys. Rev. A {\bf 78}, 033608 (2008).
\bibitem{LK08} A.~M.~L\"auchli and C.~Kollath, J.~Stat.~Mech.: Theory and Experiment, P05018 (2008).
\bibitem{BRK11} J.-S.~Bernier, G.~Roux, and C.~Kollath, Phys. Rev. Lett. {\bf 106} 200601 (2011).
\bibitem{BPBRK12} J-S.~Bernier, D.~Poletti, P.~Barmettler, G.~Roux, and C.~Kollath, Phys. Rev. A
{\bf 85}, 033641 (2012).
\bibitem{BPCK12} P.~Barmettler, D.~Poletti, M.~Cheneau, and C.~Kollath, Phys. Rev. A {\bf 85}, 053625 (2012).
%
%
%
%
%
\bibitem{QKNS14} F.~Queisser, K.~V.~Krutitsky, P.~Navez, and R.~Sch\"utzhold,
Phys.~Rev.~A {\bf 89}, 033616 (2014).
\bibitem{KNQS14} K.~V.~Krutitsky, P.~Navez, F.~Queisser, and R.~Sch\"utzhold,
EPJ Quant. Tech. {\bf1} 12 (2014).
\bibitem{NQS14} P.~Navez, F.~Queisser, and R.~Sch\"utzhold,
Jour.~Phys. A: Math. and Theor. {\bf 47} 225004 (2014).
\bibitem{QNS12} F.~Queisser, P.~Navez, and R.~Sch\"utzhold,
Phys.~Rev.~A {\bf 85}, 033625 (2012).
\bibitem{NQS16} P.~Navez, F.~Queisser, and R.~Sch\"utzhold, Phys.~Rev.~A {\bf 94}, 023629 (2016).
\bibitem{NS10} P.~Navez and R.~Sch\"utzhold, Phys.~Rev.~A {\bf 82}, 063603 (2010).
%
%
%
\bibitem{E10} T.~Esslinger, Ann.~Rev.~Cond.~Mat.~Phys. {\bf 1} 129 (2010).
\bibitem{JSGME08} R.~J\"ordens, N.~Strohmaier, K.~G\"unter, H.~Moritz, and T.~Esslinger,
Nature {\bf 455}  204 (2008).
\bibitem{CNLOZ16} L.~W.~Cheuk, M.~A.~Nichols, K.~R.~Lawrence, M.~Okan, H.~Zhang, E.~Khatami, N.~Trivedi, T.~Paiva, M.~Rigol, M.~W.~Zwierlein,
Science {\bf 353} 1260 (2016).
\bibitem{BHSON16} M.~Boll, T.~A.~Hilker, G.~Salomon, A.~Omran, J.~Nespolo, L.~Pollet, I.~Bloch, C.~Gross,
Science {\bf 353} 1257 (2016).
\bibitem{PMCJGG16} M.~F.~Parsons, A.~Mazurenko, C~S.~Chiu, Ge.~Ji, D.~Greif, M.~Greiner,
Science  {\bf 353} 1253 (2016).
%
%
Experiment Hubbard solid state
\bibitem{LNW06} P.~A.~Lee, N.~Nagaosa, and X.-G.~Wen
Rev. Mod. Phys. {\bf 78}, 17 (2006).
\bibitem{DHS03} A.~Damascelli, Z.~Hussain, and Z.-X.~Shen
Rev. Mod. Phys. {\bf 75}, 473 (2003).
%
\bibitem{MK09} M.~Moeckel and S.~Kehrein, Ann.~ Phys. {\bf 324} 2146 (2009).
%
\bibitem{SDM00} R.~Staudt, M.~Dzierzawa, and A.~Muramatsu, The European Phys.~J.~B: Cond. Mat. {\bf 17} 411 (200).


\end{thebibliography}
\end{document}